\documentclass[aps,prb,twocolumn,groupedaddress,showpacs,amsmath,fleqn]{revtex4}
\usepackage{graphicx}
\usepackage{dcolumn}
\usepackage{caption2}
\usepackage{flafter}
\usepackage{bm}
\begin{document}
\bibliographystyle{apsrev}
\title{Electric field dependence of spin coherence in (001) GaAs/AlGaAs quantum wells}
\author{Wayne H. Lau}
\altaffiliation{To whom correspondence should be addressed. E-mail: wlau@mailaps.org.}
\affiliation{Center for Spintronics and Quantum Computation, The University of California, Santa Barbara, California, 93106, USA}
\author{Michael E. Flatt\'e}
\affiliation{Department of Physics and Astronomy, The University of Iowa, Iowa City, Iowa 52242}
\date{\today}
\begin{abstract}
Conduction electron spin lifetimes ($T_1$) and spin coherence times ($T_2$) are strongly modified in semiconductor quantum wells by electric fields. Quantitative calculations in GaAs/AlGaAs quantum wells at room temperature show roughly a factor of four enhancement in the spin lifetimes at optimal values of the electric fields. The much smaller enhancement compared to previous calculations is due to overestimates of the zero-field spin lifetime and the importance of nonlinear effects.
\end{abstract}
\maketitle
The emerging field of semiconductor spintronics concerns the encoding, manipulation and detection of coherence in the spin degree of freedom of mobile electrons in semiconductors\cite{spintronics-book,sawolfscience}. Configuring semiconductor materials in quantum wells (QWs) or other artificial structures can shorten the spin lifetimes and coherence times\cite{DK} or lengthen them\cite{Ohno}. Electron spin coherence times are sensitive to carrier densities\cite{Optical-Orientation,KA1998,BaumbergPRL} and electric fields\cite{Harley2003}. A series of calculations has suggested that an electric field ($\mbox{\boldmath${\cal F}$}$) can dramatically lengthen spin lifetimes in (001) QWs for spins oriented along either the $(110)$ or $(1\overline{1}0)$ directions\cite{Averkiev-Golub, Ting}. Similar considerations have led to the proposal of coherent diffusive spin rotation in QWs\cite{LossPRL,LossAPL}.  These calculations rely on perturbative models of the spin splitting in zincblende semiconductor QWs, where the $\mbox{\boldmath${\cal F}$}$ $=$ $0$ splittings are the sum of linear and cubic terms in the electron crystal momentum ${\bf K}$ and the splitting proportional to the electric field (Rashba Hamiltonian)\cite{Rashba60,Rashba84} is also linear in ${\bf K}$.  

Here we report detailed calculations of spin lifetimes $(T_1)$ and spin coherence times $(T_2)$ in several GaAs/AlGaAs QWs at room temperature as a function of electric field. These calculations are performed using a fourteen-band electronic structure theory\cite{WHL-JTO-MEF-cm,WHL-JTO-MEF-2001} which treats the spin splitting non-perturbatively (to all orders in ${\bf K}$). We find that the Rashba spin splitting deviates from a linear dependence on ${\bf K}$ for energies within about 100 meV of the band edge. We further find that the room-temperature electric-field-induced enhancement in the spin lifetime, for QWs between 50\AA\ and 150\AA, is only a factor of four.

The Hamiltonian for bulk semiconductors in the presence of an external electric field is
\begin{equation}
\hat{H} = \frac{\hat{\bf p}^{2}}{2m_\text{e}} + V(\hat{\bf r})
+ \frac{\hbar}{4m_\text{e}^{2}c^{2}} [{\nabla} V(\hat{\bf r})  {\times}  \hat{\bf p}] \cdot \hat{\mbox{\boldmath$\sigma$}} 
+ V_{\text{ext}} (\hat{\bf r})
\label{hamCry}
\end{equation}
where $m_\text{e}$ is the free electron mass, $c$ is the velocity of light, $e$ is the elementry charge, $\hat{\bf p}$ is the electron momentum operator, $V(\hat{\bf r})$ is the periodic crystal potential, $\hat{\mbox{\boldmath$\sigma$}}$ is the Pauli spin operator, and $V_{\text{ext}} (\hat{\bf r})$ $=$ $e \mbox{\boldmath${\cal F}$} \cdot \hat{\bf r}$ is the scalar potential generated by the external electric field $\mbox{\boldmath${\cal F}$}$. The Schr$\ddot{\rm o}$dinger equation for heterostructure superlattices (SLs) is written as
\begin{equation}
\hat{H}^\text{SL} \langle {\bf r} \arrowvert {\cal L}, {\cal S}, {\bf K} \rangle 
= E_{\cal L S} ({\bf K}) \langle {\bf r} \arrowvert {\cal L}, {\cal S}, {\bf K} \rangle,
\label{sch}
\end{equation}
and the SL Hamiltonian $\hat{H}^\text{SL}$ is given by 
\begin{equation}
\hat{H}^{\text{SL}} = \sum^{N_\text{layer}}_{i = 1} \hat{H}_{i} \theta_i ({\bf r}), \quad
\theta_i ({\bf r}) =
\begin{cases}
1& \text{if ${\bf r} \in$ $i$th layer}, \\
0& \text{if ${\bf r} \not \in$ $i$th layer},
\end{cases}
\label{hamSL}
\end{equation}
where $\hat{H}_{i}$ is the crystal Hamiltonian of the $i$th layer [Eq.~(\ref{hamCry})] with parameters tabulated in Ref.~\onlinecite{WHL-JTO-MEF-cm}, $N_\text{layer}$ is the number of layers in the SL unit cell, $\arrowvert {\cal L},  {\cal S}, {\bf K} \rangle$ is the SL eigenstate for a carrier with wavevector ${\bf K}$, pseudo spin quantum number ${\cal S}$ $\in$ $\{\uparrow, \downarrow\}$, and band index ${\cal L}$, and $E_{\cal L S} ( {\bf K})$ is the corresponding SL eigenenergy. To obtain  $\arrowvert {\cal L}, {\cal S}, {\bf K} \rangle$ and $E_{\cal L S} ({\bf K})$, we first solve the Schr$\ddot{\rm o}$dinger equation [Eq.~(\ref{sch})] at the zone center in which the SL wavevector ${\bf K}$ $=$ ${\bf 0}$. The solution, 
\begin{equation}
 \langle {\bf r} \arrowvert N, S, {\bf 0} \rangle 
 = \sum_{n \sigma} F_{N S n \sigma} ({\bf r}) \langle {\bf r} \arrowvert n, \sigma, {\bf 0} \rangle, 
\label{zcwf}
\end{equation}
where $\arrowvert n, \sigma, {\bf 0} \rangle$ are the corresponding zone-center Bloch states of the constituent bulk semiconductors and the expansion coefficients $F_{N S n \sigma} ({\bf r})$ are slowly varying envelope functions on the scale of the constituent bulk semiconductor lattice constant. The zone-center SL wavefunctions and energies are found as described in Ref.~\onlinecite{WHL-JTO-MEF-cm},  Sec.~IIIB.

Once the zone-center SL eigenstates and eigenenergies have been determinted, the SL states for ${\bf K} \neq {\bf 0}$ can be obtained by application of a generalized SL ${\bf K} \cdot {\bf p}$ theory. A SL state $\arrowvert {\cal L}, {\cal S}, {\bf K} \rangle$ at finite ${\bf K}$ is expressed in terms of the zone-center SL states as $\langle {\bf r} \arrowvert {\cal L}, {\cal S}, {\bf K} \rangle$ 
$=$ $\exp(i {\bf K \cdot r}) \sum_{N S} C_{{\cal L S} N S} ({\bf K}) \langle {\bf r} \arrowvert N, S, {\bf 0} \rangle$, where $C_{{\cal L S} N S} ({\bf K})$ are the expansion coefficients. By inserting $\langle {\bf r} \arrowvert {\cal L}, {\cal S}, {\bf K} \rangle$ into Eq.~(\ref{sch}), multipling the resulting equation on the left-hand side by $\exp({\bf -K \cdot r}) \langle N, S, {\bf 0} \arrowvert {\bf r} \rangle$ and integrating over a SL unit cell, we have
\begin{eqnarray} 
&&\sum_{N' S'} \bigg[\left(\frac{\hbar^2 {\bf K}^2}{2m_{e}} 
+ E_{N S} ({\bf 0}) - E_{\cal L S} ({\bf K}) \right) \delta_{N N'} \delta_{S S'} \nonumber \\
&&+ \frac{\hbar}{m_{e}}{\bf K} \cdot {\bf P}_{N S N' S'} ({\bf 0}) \bigg]
C_{{\cal L S} N' S'} ({\bf K}) =  0, 
\label{sse}
\end{eqnarray} 
where ${\bf P}_{N S N' S'} ({\bf 0})$ $\equiv$ $\langle N, S, {\bf 0} \arrowvert \hat{{\bf p}} \arrowvert N', S', {\bf 0} \rangle$.

Spin splitting in zincblende QW's can be described with an effective momentum-dependent internal magnetic field, ${\bf H}({\cal L}, {\bf K}) $ $=$ $\Upsilon^{-1/2} {\bf \Omega} ({\cal L}, {\bf K})$, where $\Upsilon \equiv (\text{g} \mu_{\text{B}} \hbar^{-1})^2$, $\mu_{\text{B}}$ is the Bohr magneton, and $\text{g}$ is the electron g-factor. The components of the spin precession vector ${\bf \Omega} ({\cal L}, {\bf K})$ are given by the following relations\cite{WHL-JTO-MEF-cm}:
\begin{subequations}
\label{om}
\begin{equation}
\Omega_{x}({\cal L}, {\bf K})
= \text{Re} \left[ 2 \Omega \sum_{N} C_{{\cal L} \downarrow N \uparrow}^{\ast} ({\bf K}) 
C_{{\cal L} \uparrow N \uparrow} ({\bf K}) \right],
\label{omx}
\end{equation}
\begin{equation}
\Omega_{y}({\cal L}, {\bf K})
= \text{Im} \left[ 2 \Omega \sum_{N} C_{{\cal L} \downarrow N \uparrow}^{\ast} ({\bf K}) 
C_{{\cal L} \uparrow N \uparrow} ({\bf K}) \right],
\label{omy}
\end{equation}
\begin{equation}
\Omega_{z}({\cal L}, {\bf K})
= \sum_{N} \Omega \Big[ \left| C_{{\cal L} \uparrow N \uparrow} ({\bf K}) \right|^{2} 
- \left[ C_{{\cal L} \downarrow N \uparrow} ({\bf K}) \right|^{2} \Big],
\label{omz}
\end{equation}
\end{subequations}
where the magnitude of the spin precession vector is given by $\Omega$ $=$ $\hbar^{-1}| E_{{\cal L} \uparrow} ({\bf K}) - E_{{\cal L} \downarrow} ({\bf K})|$. 

This spin splitting is due to the combined effects of spin-orbit interaction and spatial inversion asymmetry, where this spatial inversion asymmetry can arise from the bulk inversion asymmetry (BIA) of the constituent semiconductors and the structural inversion asymmetry (SIA) of the QWs. The spin precession vector can be decomposed into ${\bf \Omega} ({\cal L}, {\bf K})$ $=$ ${\bf \Omega}^\text{(D)}({\cal L}, {\bf K})$ $+$ ${\bf\Omega}^\text{(R)}({\cal L}, {\bf K})$ according to its symmetry, where ${\bf \Omega}^{\text{(D)}}({\cal L}, {\bf K})$ and ${\bf \Omega}^{\text{(R)}}({\cal L}, {\bf K})$ are the momentum-dependent spin precession vectors due to bulk inversion asymmetry and structural inversion asymmetry, respectively. Although many analyses of these spin splitting fields focus on the ${\bf K}$-linear terms, here we keep terms to all orders.

In the linear approximations, both $[{\bf \Omega}_{1}^\text{(D)}({\cal L}, E)]^2$ and $[{\bf \Omega}_{1}^\text{(R)}({\cal L}, E)]^2$ are assumed proportional to $E$ for all energies~\cite{definition-fourier-component}. In addition, the nonlinear $[{\bf \Omega}_{3}^\text{(D)}({\cal L}, E)]$ and $[{\bf \Omega}_{3}^\text{(R)}({\cal L}, E)]$, are neglected in other analyses. Depending on the size of the quantum well, these two terms can become dominant. The importance of these nonlinear effects was pointed out in symmetric quantum wells~\cite{WHL-JTO-MEF-2001}. Multiband calculations for asymmetric quantum wells have been done, but as a function of asymmetric doping, not as a function of an applied electric field~\cite{winkler98, winkler03}.
\begin{figure}[htbp]
\setcaptionwidth{8.5 cm}
\scalebox{1}{\includegraphics[width = 8.5 cm]{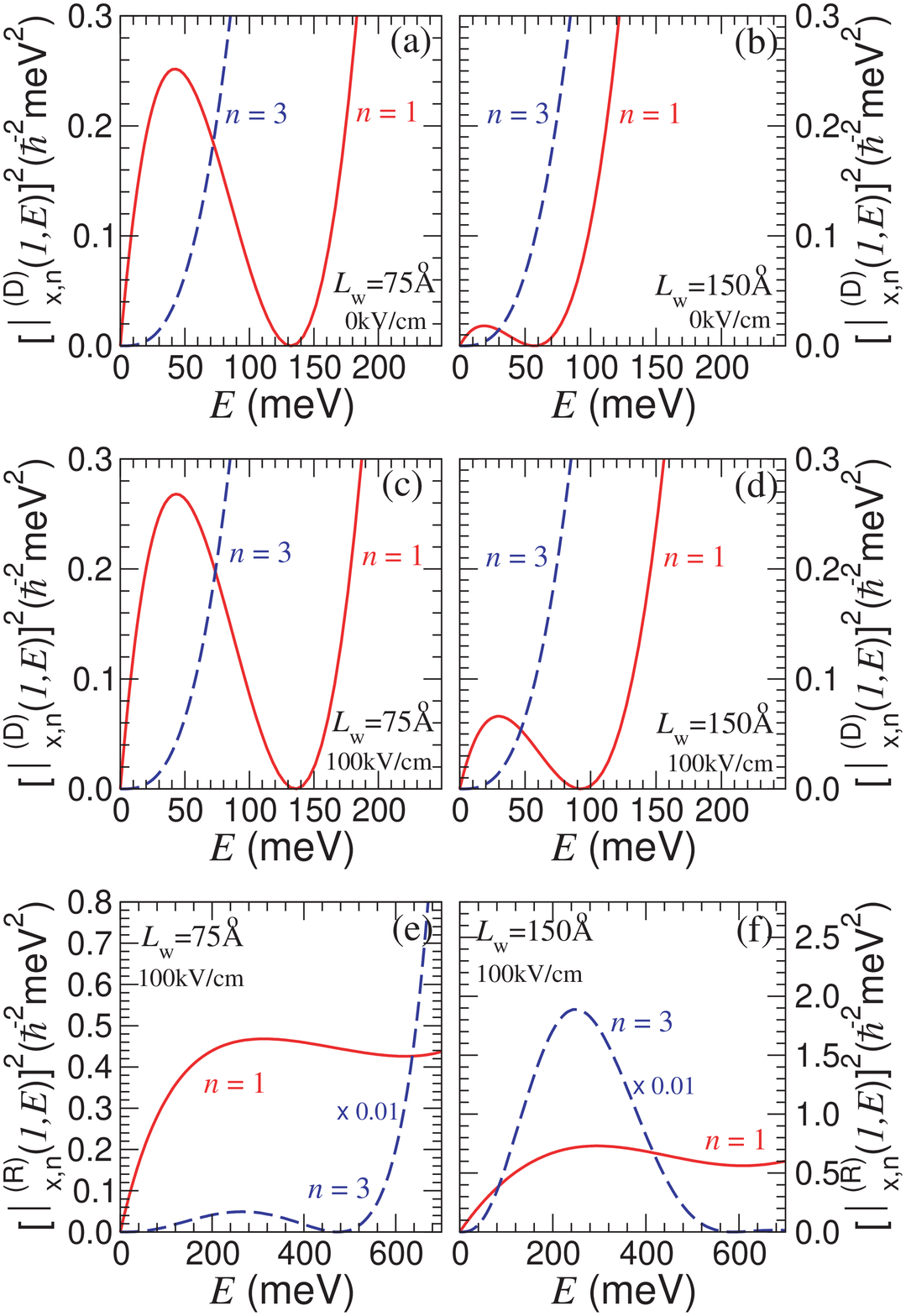}}
\caption[]{(color online). Energy dependence of electron spin precession vector for a $75$-\AA\ and a $150$-\AA\ GaAs/$100$-\AA\ Ga$_{0.6}$Al$_{0.4}$As QW at $300$K. $[\Omega^\text{(D)}_{\text{x}, n}({1, E})]^2$ as a function of $E$ for a $75$-\AA\ QW (a) and a $150$-\AA\ QW (b) with $\cal F$ $=$ $0$ kV/cm. $[\Omega^\text{(D)}_{\text{x}, n}({1, E})]^2$ as a function of $E$ for a $75$-\AA\ QW (c) and a $150$-\AA\ QW (d) with $\cal F$ $=$ $100$ kV/cm. $[\Omega^\text{(R)}_{\text{x}, n}({1, E})]^2$ as a function of $E$ for a $75$-\AA\ QW (e) and a $150$-\AA\ QW (f) with $\cal F$ $=$ $100$ kV/cm.}
\label{GaAsAlGaAsOL}
\end{figure}

Electron spin relaxation near room temperature is dominated by the D'yakonov-Perel' (DP) mechanism due to the effective internal magnetic field arises from spin-orbit interaction\cite{DP-mech, DK, Optical-Orientation}. In the motional narrowing regime the electronic spin system is subject to a {\it time-dependent}, randomly oriented effective internal magnetic field ${\bf H}({\cal L}, E)$ which changes direction with an orbital scattering time $\tau({\cal L}, E)$ that is much shorter than the precession time  of either the constant applied magnetic field ${\bf H_o}$ or the random field. The spin relaxation time and spin coherence time in zincblende QWs depend on the transverse $H_\perp ({\cal L}, E)$ and longitudinal $H_\parallel ({\cal L}, E)$ components of the random field, according to\begin{subequations}
\label{t1t2}
\begin{eqnarray} 
T_1^{-1}  
= \Upsilon \sum_{\cal L} \int d E {\cal D}({\cal L}, E) H_\perp^2({\cal L}, E) \tau({\cal L}, E)  h({\cal L}, E),
\label{t1}
\end{eqnarray}
\begin{eqnarray} 
T_2^{-1}  
&=&  \Upsilon  \sum_{\cal L} \int d E {\cal D}({\cal L}, E)
\left[ \frac{1}{2}H_\perp^2({\cal L}, E) + H_\parallel^2({\cal L}, E) \right]  \nonumber \\ 
&\times& \tau({\cal L}, E) h({\cal L}, E),
\label{t2}
\end{eqnarray}
\end{subequations} 
where ${\cal D}({\cal L}, E)$ is the density of states and $h({\cal L}, E)$ is the electron distribution function\cite{WHL-JTO-MEF-cm}.

Now we present the results of our quantitative numerical calculations of the spin precession vectors, spin lifetimes, and spin coherence times of (001) GaAs/AlGaAs quantum wells. We find that the nonlinear effects are important and both $[{\bf \Omega}_{3}^\text{(D)}({\cal L}, E)]$ and $[{\bf \Omega}_{3}^\text{(R)}({\cal L}, E)]$are significant in these quantum wells. In the absence of an external electric field, the effective internal magnetic field in a perfectly symmetric QW arises entirely from the BIA terms, and the SIA terms vanish identically. Figures~\ref{GaAsAlGaAsOL}(a) and~\ref{GaAsAlGaAsOL}(b) show $[\Omega^\text{(D)}_{\text{x}, n}({1, E})]^2$ as a function of energy $E$ calculated using Eqs.~(\ref{om}) for a $75$-\AA\ and a $150$-\AA\ QW, respectively. $[\Omega^\text{(D)}_{\text{x}, 1}({1, E})]^2$ is proportional to $E$ and $E^3$, while $[\Omega^\text{(D)}_{\text{x}, 3}({1, E})]^2$ is proportional to $E^3$, consistent with the analytical results\cite{DK,Rashba60,Rashba84}. $[\Omega^\text{(D)}_{\text{x}, n}({1, E})]^2$ decreases as the width $L_\text{w}$ of the QW increases; and it deviates from a linear dependence on $E$ within $50$ meV for QWs between $50$\AA\ and $150$\AA\ indicating the importance of the nonlinear effects. 

When an external electric field is applied along the growth direction ($\mbox{\boldmath${\cal F}$} \| \hat{\bf z})$, the breaking of structural inversion symmetry results in modification of the effective internal magnetic field through both the BIA and SIA terms [Figs.~\ref{GaAsAlGaAsOL}(c)-\ref{GaAsAlGaAsOL}(f)]. $[\Omega^\text{(D)}_{\text{x}, 1}({1, E})]^2$ increases as ${\cal F}$ increases, and it also increases with increasing $L_\text{w}$ [cf., Figs.~\ref{GaAsAlGaAsOL}(a)-\ref{GaAsAlGaAsOL}(d)]. For the energy range of physical interests, $[\Omega^{\text{(R)}}_{x,1}(1, E)]^2$ is approximately two order in magnitude larger than $[\Omega^{\text{(R)}}_{x, 3}(1, E)]^2$; and $[\Omega^{\text{(R)}}_{x,1}(1, E)]^2$ deviates from a linear dependence on $E$ for energies within $100$ meV [shown in Figs.~\ref{GaAsAlGaAsOL}(e) and~\ref{GaAsAlGaAsOL}(f)].
\begin{figure}[htbp]
\setcaptionwidth{8.5 cm}
\scalebox{1}{\includegraphics[width = 8.5 cm]{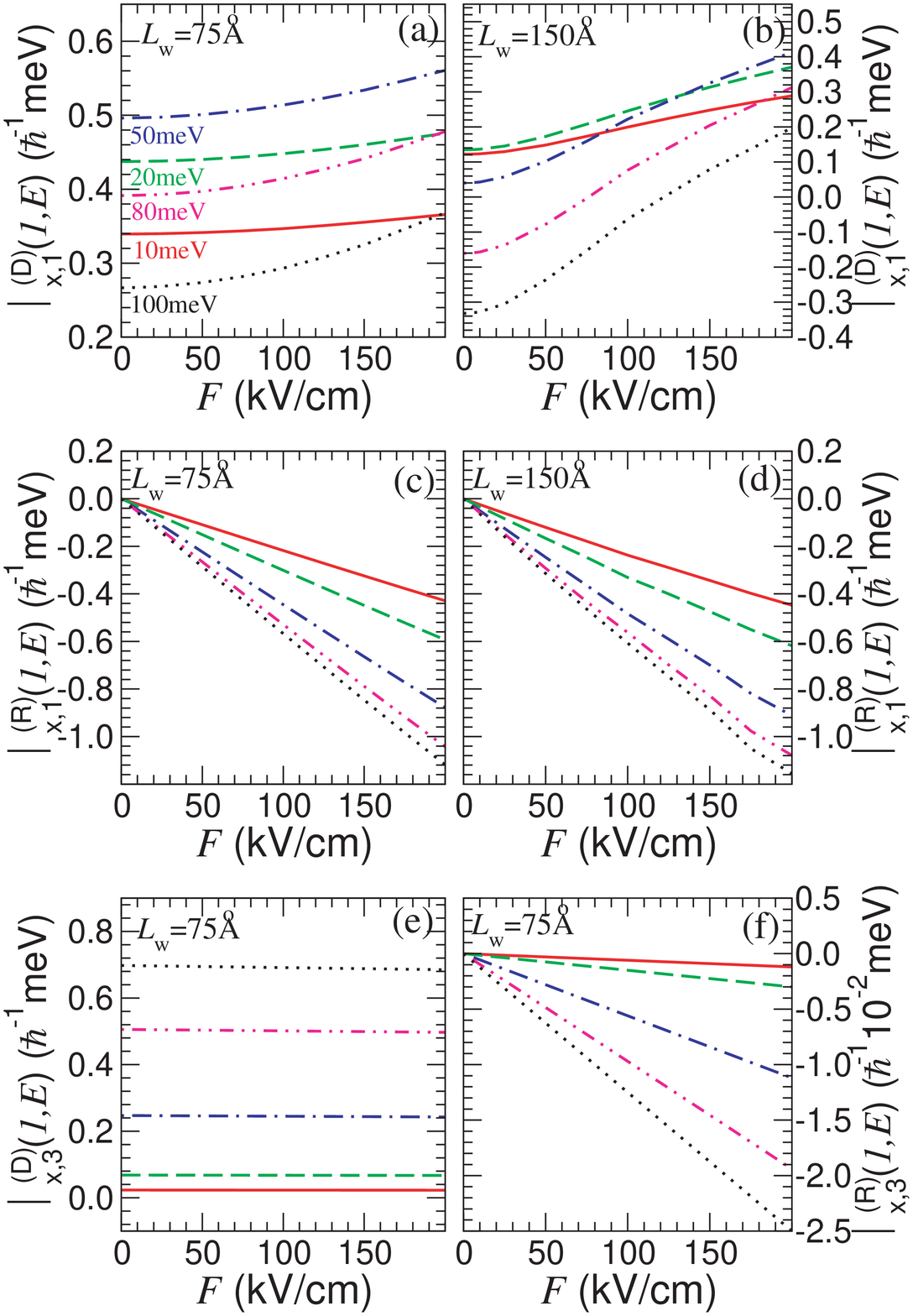}}
\caption[]{(color online). Electric field dependence of electron spin precession vector for a $75$-\AA\ and a $150$-\AA\ GaAs/$100$-\AA\ Ga$_{0.6}$Al$_{0.4}$As QW at $300$K. $\Omega^\text{(D)}_{\text{x}, 1}({1,E})$ as a function of $\cal F$ for a $75$-\AA\ QW (a) and a $150$-\AA\ QW (b). $\Omega^\text{(R)}_{\text{x}, 1}({1,E})$ as a function of $\cal F$ for a $75$-\AA\ QW (c) and a $150$-\AA\ QW (d). (e) $\Omega^\text{(D)}_{\text{x}, 3}$ as a function of $\cal F$ for a $75$-\AA\ QW. (f) $\Omega^\text{(R)}_{\text{x}, 3}$ as a function of $\cal F$ for a $75$-\AA\ QW.}
\label{GaAsAlGaAsOF}
\end{figure}

We find that the electric field dependence of $\Omega^\text{(R)}_{\text{x}, 1}({1,E})$ is linear, whereas that of $\Omega^\text{(D)}_{\text{x}, 1}({1,E})$ is nonlinear. The electric field dependence of $\Omega^\text{(D)}_{\text{x}, 3}({1,E})$ is much weaker compared to that of $\Omega^\text{(R)}_{\text{x}, 3}({1,E})$. To explore the electric field dependence of the effective internal magnetic field, $\Omega^\text{(D)}_{\text{x}, n}({1,E})$ and $\Omega^\text{(R)}_{\text{x}, n}({1,E})$ as a function of ${\cal F}$ are calculated for various values of $E$ and the results for a $75$-\AA\ and a $150$-\AA\ QW are shown in Figs.~\ref{GaAsAlGaAsOF}(a)-\ref{GaAsAlGaAsOF}(f). For a constant value of $E$, the magnitude of $\Omega^\text{(D)}_{\text{x}, 1}({1,E})$ and $\Omega^\text{(R)}_{\text{x}, 1}({1,E})$ increases with increasing ${\cal F}$; and the slope of the curves is positive for the BIA term, while it is negative for the SIA term. The nonlinear effects are also manifested in the electric field dependence of $\Omega^\text{(D)}_{\text{x}, 1}({1,E})$ [see Figs.~\ref{GaAsAlGaAsOF}(a) and \ref{GaAsAlGaAsOF}(b)].
\begin{figure}[htbp]
\setcaptionwidth{8.5 cm}
\scalebox{1}{\includegraphics[width = 8.5 cm]{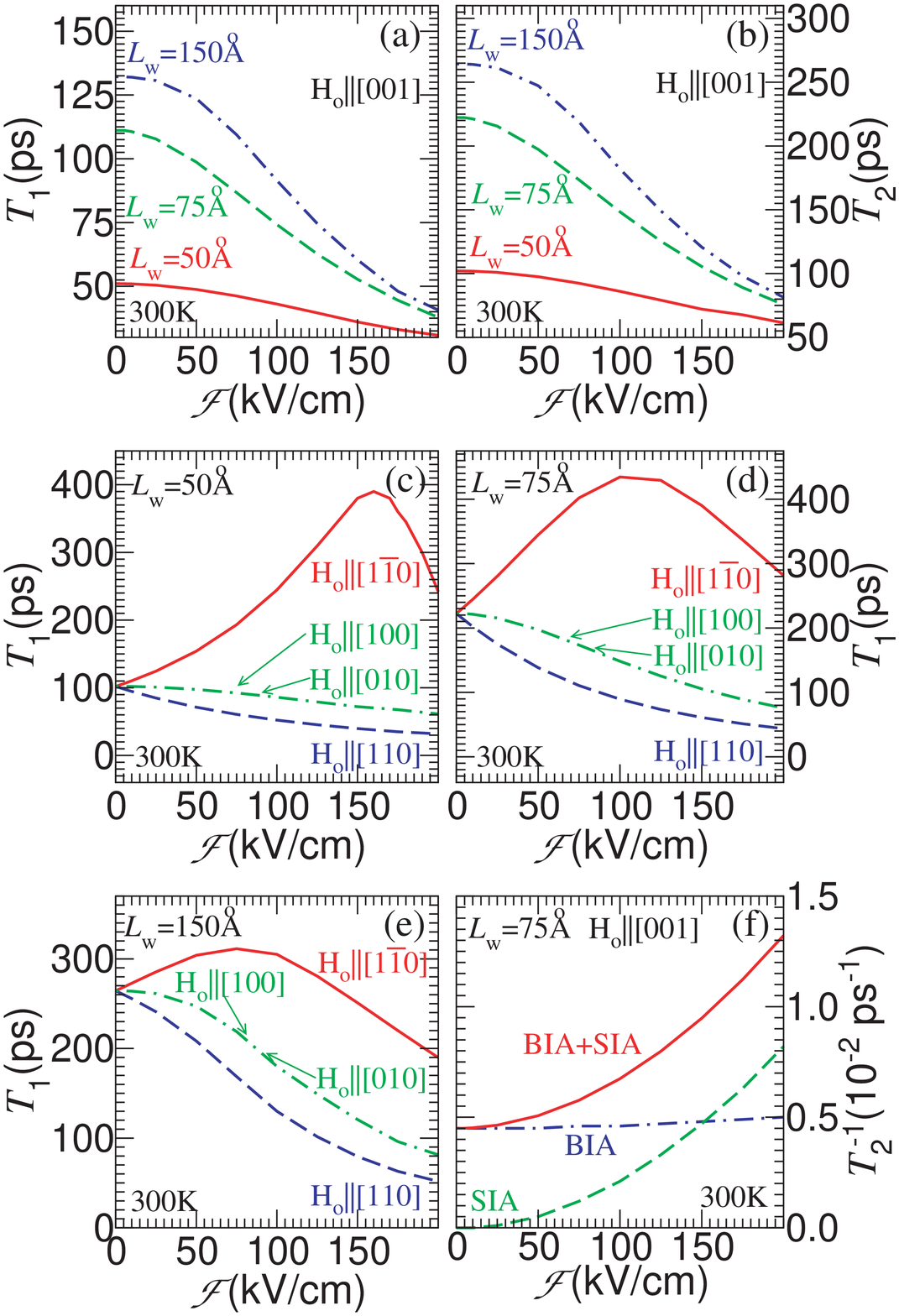}}
\caption[]{(color online). Electron spin lifetimes, coherence times, and decoherence rates as a function of applied electric field $\cal F$ for a $50$-\AA, a $75$-\AA, and a $150$-\AA\ GaAs/$100$-\AA Ga$_{0.6}$Al$_{0.4}$As QWs at $300$K with $\mu$ $=$ $800$ cm$^2$/Vs assuming neutral impurity scattering. (a) $T_1$ as a function of $\cal F$ with ${\bf H}_\text{o}||[001]$.  (b) $T_2$ as a function of $\cal F$ with ${\bf H}_\text{o}||[001]$.  (c) $T_1$ as a function of ${\cal \cal F}$ with ${\bf H}_\text{o}||[110]$, ${\bf H}_\text{o}||[1\overline{1}0]$, ${\bf H}_\text{o}||[100]$, and ${\bf H}_\text{o}||[010]$ for a $50$-\AA\ QW. (d) $T_1$ as a function of ${\cal \cal F}$ with ${\bf H}_\text{o}||[110]$, ${\bf H}_\text{o}||[1\overline{1}0]$, ${\bf H}_\text{o}||[100]$, and ${\bf H}_\text{o}||[010]$ for a $75$-\AA\ QW. (e) $T_1$ as a function of ${\cal \cal F}$ with ${\bf H}_\text{o}||[110]$, ${\bf H}_\text{o}||[1\overline{1}0]$, ${\bf H}_\text{o}||[1\overline{1}0]$, and ${\bf H}_\text{o}||[1\overline{1}0]$ for a $150$-\AA\ QW. (f) $T^{-1}_2$ as a function of $\cal F$ with ${\bf H}_\text{o}||[001]$.}
\label{GaAsAlGaAsT12F}
\end{figure}

Both $T_1$ and $T_2$ are more responsive to the electric field for wide QWs [i.e., $(dT_{1,2}/d{\cal F})$ increases as ${L_\text{w}}$ increases], and this can be seen by examining the change in $\Omega_{\text{x}, n}({1,E})$ with ${\cal F}$ [cf., Figs.~\ref{GaAsAlGaAsOL}(a)-\ref{GaAsAlGaAsOL}(d)]. For example, the change in $\Omega^\text{(D)}_{\text{x}, 1}({1,E})$ from zero field to $100$ kV/cm is much larger for the $150$-\AA\ QW than for the $75$-\AA\ QW. The calculations of electron spin lifetime and electron spin coherence time are performed using Eqs.~(\ref{t1t2}). To examine the electric field dependence of spin lifetime and spin coherence time with applied magnetic field along the growth direction, $T_1$ and $T_2$ as a function of ${\cal F}$ are calculated for a $50$-\AA, a $75$-\AA, and a $150$-\AA\ QWs at 300K, and the results are shown in Figs.~\ref{GaAsAlGaAsT12F}(a) and \ref{GaAsAlGaAsT12F}(b), respectively. For the DP mechanism, the spin lifetime and spin coherence time are inversely proportional to the square of the effective internal magnetic field\cite{DP-mech}; and consequently both the spin lifetime and the spin coherence time decrease with increasing electric field due to the increase of the total effective internal magnetic field with electric field. 

We find that the spin-lifetime enhancement originates from the destructive interference between the BIA and SIA effective internal magnetic fields due to the symmetry breaking of structural inversion symmetry in the presence of an electric field. We also find that the calculated enhancement of the spin lifetime is much smaller than that of the previous calculations based on perturbative approaches\cite{Ting}, in which the zero-field spin lifetime is overestimated and the nonlinear effects are not taken into account. To investigate the electric-field-induced enhancement in the electron spin lifetime\cite{Averkiev-Golub}, $T_1$ as a function of ${\cal F}$ for three QWs of different thickness is calculated for an applied magnetic field parallel to the in-plane direction and the results are shown in Figs.~\ref{GaAsAlGaAsT12F}(c)-\ref{GaAsAlGaAsT12F}(e). As expected the calculated spin lifetime is identical for ${\bf H}_\text{o}||[100]$ and ${\bf H}_\text{o}||[010]$ due to the crystal symmetry of $(001)$ QWs, and $T_1$ decreases with increasing ${\cal F}$. The electric field dependence of $T_1$ for ${\bf H}_\text{o}||[110]$ is similar to that for ${\bf H}_\text{o}||[100]$, and the spin lifetime is shorter for ${\bf H}_\text{o}||[110]$ than for ${\bf H}_\text{o}||[100]$. For ${\bf H}_\text{o}||[1\overline{1}0]$, $T_1$ increases as ${\cal F}$ increases, reaching a maximum value and then it decreases as ${\cal F}$ further increases. The spin-lifetime enhancement factor [$T_1({\cal F})/T_1(0)$] decreases as ${L_\text{w}}$ increases, and the maximum enhancement factor for a $50$-\AA, a $75$-\AA, and a $150$-\AA\ QWs is approximately $3.8$, $2.0$, and $1.2$, respectively. 

In the presence of an electric field, we find that spin decoherence depends on both the BIA and SIA effective internal magnetic fields, and the spin decoherence is dominated by the BIA terms at low field while it is dominated by the SIA terms at high field. To study the effects of the BIA and the SIA effective internal magnetic fields on spin decoherence (as opposed to spin lifetimes), the decoherence rates ($T^{-1}_2$) as a function of ${\cal F}$ for a $75$-\AA\ QW due to BIA and SIA are calculated separately, and the results as well as the total decoherence rate are plotted in Fig.~\ref{GaAsAlGaAsT12F}(f). It can be seen that the spin decoherence is dominated by the BIA effective internal magnetic field for an electric field as high as $100$ kV/cm, and crossover occurs at approximately $150$ kV/cm. The value of the crossover decreases as ${L_\text{w}}$ increases, and the crossover for a $50$-\AA\ and a $150$-\AA\ QWs occurs at approximately $250$ kV/cm and $120$ kV/cm, respectively (not shown).

In summary, we have studied quantitatively the electric field dependence of electron spin coherence in $(001)$ GaAs/AlGaAs QWs using a non-pertubative fourteen-band electronic structure theory. We find that both the BIA and SIA effective internal magnetic fields are electric field dependent. The electron spin lifetime and spin coherence time are strongly influenced by an electric field. Even for moderate ($<100$ kV/cm) electric fields, the spin decoherence is dominated by the BIA effective internal magnetic field. At optimal electric fields, the enhancement factor of spin lifetime for GaAs/AlGaAs QWs between $50$-\AA\ and $150$-\AA \ is approximately 4, which is much smaller than that obtained from the previous calculations based on perturbative, linear approximations. 

This work was supported by DARPA/ARO.

\end{document}